# Last performances improvement of the C-RED One camera using the 320x256 e-APD infrared Saphira detector

Philippe Feautrier [a] and Jean-Luc Gach [a]

[a] First Light Imaging SAS, Europarc Sainte Victoire, Bâtiment 5, Route de Valbrillant, Le Canet, 13590 Meyreuil, France

## ABSTRACT

We present here the latest results obtained with the C-RED One camera developed by First Light Imaging for fast ultra-low noise infrared applications. This camera uses the Leonardo Saphira e-APD 320x256 infrared sensor in an autonomous cryogenic environment with a low vibration pulse tube and with embedded readout electronics system. Some recent improvements were made to the camera. The first important one concerns the total noise of the camera. Limited to 1.75 µm wavelength cut-off with proper cold filters, looking at a blackbody at room temperature and f/4 beam aperture, we now measure total noise down to 0.6 e at gain 50 in CDS mode 1720 FPS, dividing previous noise figure by a factor 2. The total camera background of 30-400 e/s is now achieved with a factor 3 of background reduction, the camera also looking at a room temperature blackbody with an F/4 beam aperture. Image bias oscillations, due to electronics grounding scheme, were carefully analyzed and removed. Focal plane detector vibrations transmitted by the pulse tube cooling machine were also analyzed, damped and measured down to 0.3 µm RMS, reducing focal plane vibrations by a factor 3. In addition, a vacuum getter of higher capacity is now used to offer camera operation without camera pumping during months. We also delivered C-RED One cameras with K band limitation instead of our classical H band configuration. The camera can deal with room temperature observation in K band by limiting the beam aperture to f/20. The camera main characteristics are detailed: pulse tube cooling at 80K with limited vibrations, permanent vacuum solution, ultra-low latency Cameralink full data interface, safety management of the camera by firmware, online firmware update, ambient liquid cooling and reduced weight of 20 kg. C-RED one is the only modern infrared camera offering such level of performances in terms of reliability, noise and speed.

**Keywords:** SWIR, infrared camera, high speed, C-RED One, HgCdTe, linear mode APD, Saphira, sub-e noise

## 1. INTRODUCTION

### 1.1 The Saphira Detector and C-RED One camera

C-RED1 is an ultra-low noise infrared camera based on the Saphira detector and fabricated by First Light Imaging, specialized in fast imaging camera, after the successful commercialization of the OCAM2 camera [1] dedicated to extreme adaptive optics wavefront sensing. Designed and fabricated by Leonardo UK, formerly Selex, the Saphira detector is designed for high speed infrared applications and is the result of a development program alongside the European Southern Observatory on sensors for astronomical instruments [2], [3], [4]. It delivers world leading photon sensitivity of <1 photon rms with Fowler sampling and high-speed non-destructive readout (>10K frame/s). Saphira is an HgCdTe avalanche photodiode (APD) array incorporating a full custom ROIC for applications in the 1 to 2.5µm range. C-RED One camera is an autonomous plug-and-play system with a user-friendly interface, which can be operated in extreme and remote locations. The sensor is placed in a sealed vacuum environment and cooled down to cryogenic temperature using an integrated pulse tube. The vacuum is self-managed by the camera and no human intervention is required.

The system shown in Figure 1 has been extensively described by Feautrier et al. in 2017 [5].

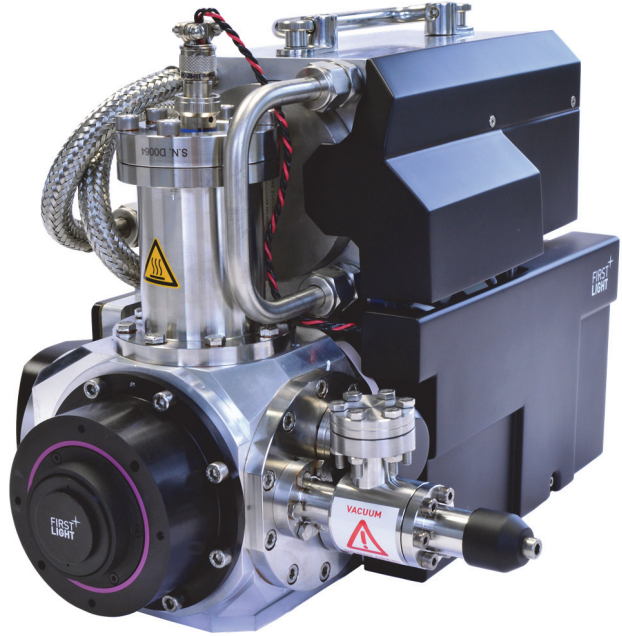

*Figure 1 : the C-RED One camera. The cooling system (pulse tube) can be seen on the top whereas in the bottom are the vacuum cryostat and the readout electronics.*

## 2. MEASURED C-RED ONE PERFORMANCES

All data reported here were recorded at a temperature of 80K.

### 2.1 Quantum efficiency

The array quantum efficiency peaks up to near 80% and the array AR coating may be optimized for J, H or K bands (H band is the standard one). Figure 2 shows the effect of this QE optimization. Moreover due to junction heterostructure with 3.5 µm cutoff wavelength HgCdTe material for the avalanche multiplication region and 2.5 µm material for the absorber, the device is sensitive in L band at gain 1 but less with APD gain. This is due to photon penetration depth (longer wavelength photons penetrate deeper in the material and therefore are less amplified).

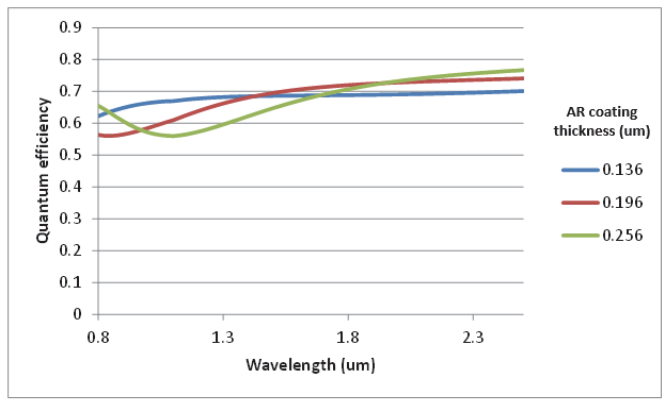

*Figure 2 : AR coating and QE optimization for J, H or K bands of Mark13 e-APD diodes. The H band coating is the standard one.*

## 2.2 Conversion gain of the camera

The Saphira transfer function with $C_{node}$= 30 fF total (27 fF diode + 3fF input amplifier) is:

$$G_1 = 0.8 \frac{e}{C_{node}} = 0.8 \, x \, \frac{1.6 \, 10^{-19}}{30 \, 10^{-15}} = 4.27 \, \mu V/e$$

The analog chain gain is: $G_2 = 10$

A/D converter conversion gain is $G_3 = \frac{4 \, V}{2^{16}} = 61 \, \mu V/adu$

The theorical conversion gain is $G = \frac{G_3}{G_1 G_2} = 1.42 \, e/adu$

The conversion gain can also be measured illuminating the sensor with a flat field through an integrating sphere. Flat fields of 2000 images are acquired with different illumination in CDS mode. For each flat field, the mean of the signal and the variance of the signal is computed. The mean variance is plotted as a function of the mean signal. The system gain can be computed as the inverse of the slope from a linear regression of the plot: this is the well-known photon transfer curve method (PTC).

Here we measure a conversion gain of 1.32 e/adu, very close to the theorical value. We can note that the PTC plot is in fact rounded and not fully linear. This is due to the pixel architecture which is not fully linear. This means that in fact the conversion gain is not constant during the integration: during the integration, the diode capacitance varies, the conversion gain varies and even the multiplication gain (when the gain is applied) varies also. The measurement gives only a mean value of the conversion gain.

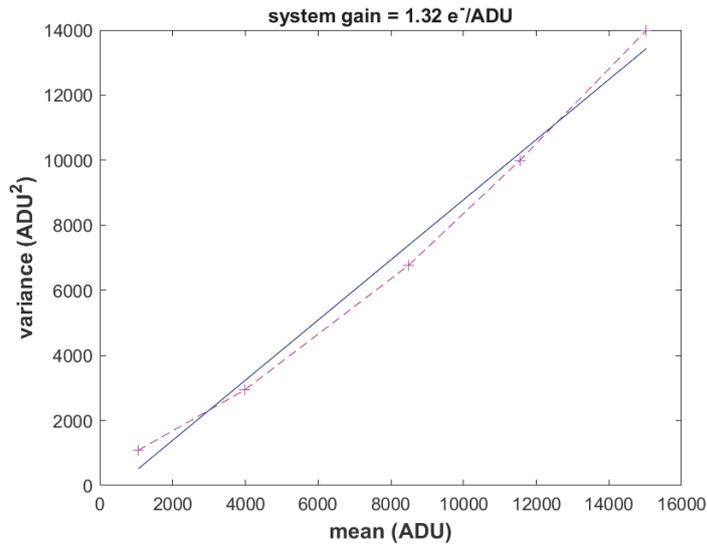

Figure 3 : Photon transfer curve of single readout mode (left) and CDS mode (right).

## 2.3 APD gain

The avalanche photodiode gain is difficult to measure directly because of the cut-off wavelength which varies with the multiplication gain. This can be seen as a quantum efficiency (QE) variation with the multiplication gain. To turn around this difficulty, we measured the multiplication gain using an IR 1300nm photodiode to maximize the photon flux below 1.8 µm where the QE does not vary with the wavelength. The multiplication gain is measured by direct division of the signal at gain M by the signal at gain 1. In practical, we are using the global reset non-destructive readout mode with 3 readouts and we simply divide the signal slopes between the 2 cases using signals acquired exactly in the same conditions. This is then a direct measurement of the gain, with no assumption on the noise characteristics. We only must assume that the Saphira QE in the spectral bandwidth of the IR photo diode does not change when the multiplication gain is applied. Once the multiplication gain is known as a function of the APD diode bias voltage, the curve is fitted using an exponential fit (see **Erreur ! Source du renvoi introuvable.**). This allows to build a multiplication gain table which is loaded in the camera.

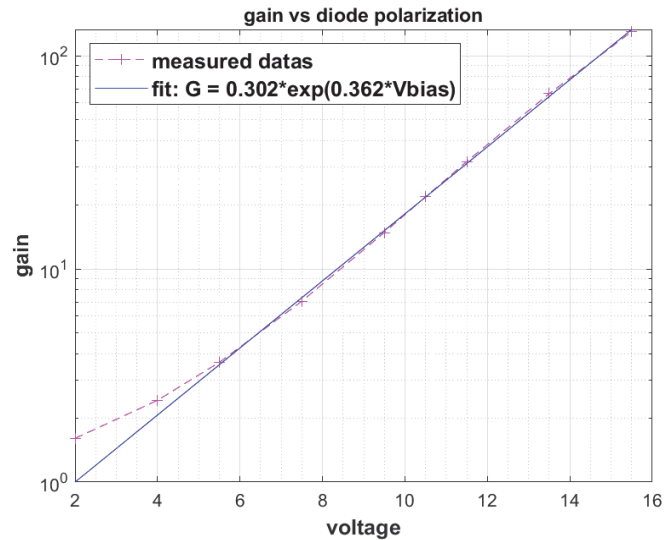

*Figure 4 : Measured APD gain vs polarization voltage of the Saphira array and exponential fit.*

## 2.4 Camera noise

The noise measurement is done by measuring the temporal variation of the image, sensor in the dark running at 1720fps.

Conditions of set-up:

- Detector temperature: 80K
- Read-out mode: CDS full frame 1720 fps
- eAPD gain: 50
- Cold stop in front of the detector (inside the cryostat); cold baffle and filter are not mounted
- Water temperature: 20°C.

2000 frames is collected. The read-out noise measurement is performed by computing the temporal standard deviation of the pixel values over the full frame of the detector (statistics on the 2000 images of the data cube, the mean readout noise of the full frame is computed

Noise results are shown hereafter in the Figure 5:

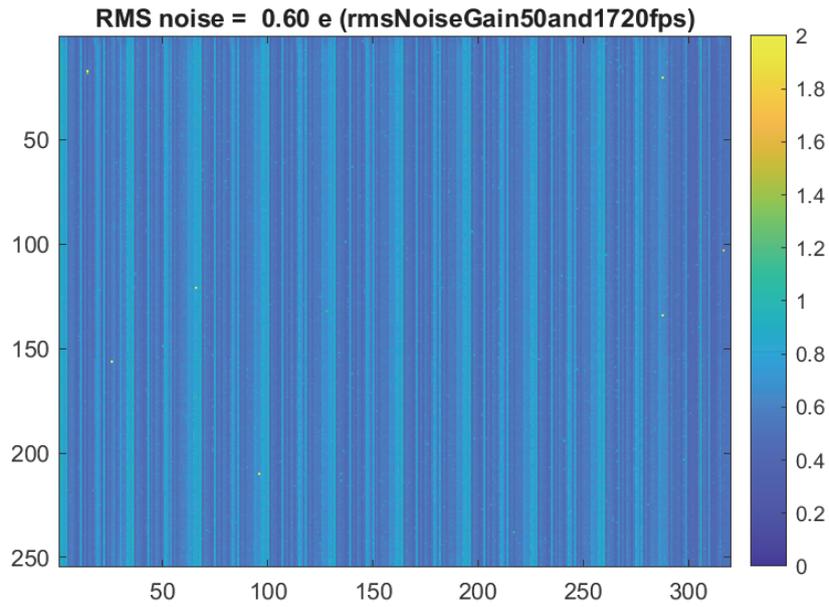

*Figure 5 : the mean RMS noise at Gain = 50 and 1720 fps*

The RMS Noise with a cold stop in front of the detector at Gain x50 in CDS mode, 1720 fps is **0.60 e-.**

When looking at a room temperature blank placed in front of the camera window, the noise increases slightly: 0.67 e CDS mode, gain 50 and 1720 fps (see Figure 6), 0.67 e CDS mode, gain 50 and 1000 fps (see Figure 7).

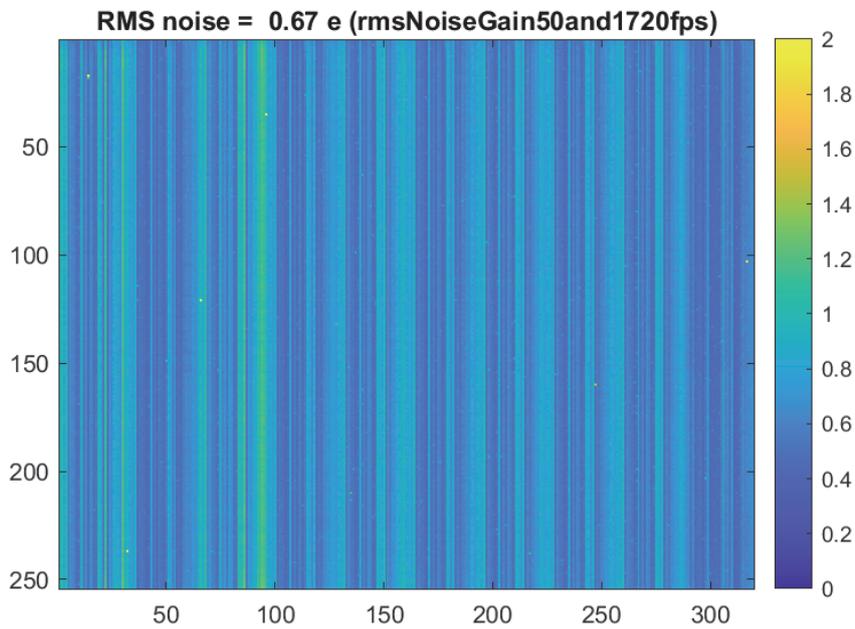

*Figure 6: CDS noise, Gain x50, 1720 fps*

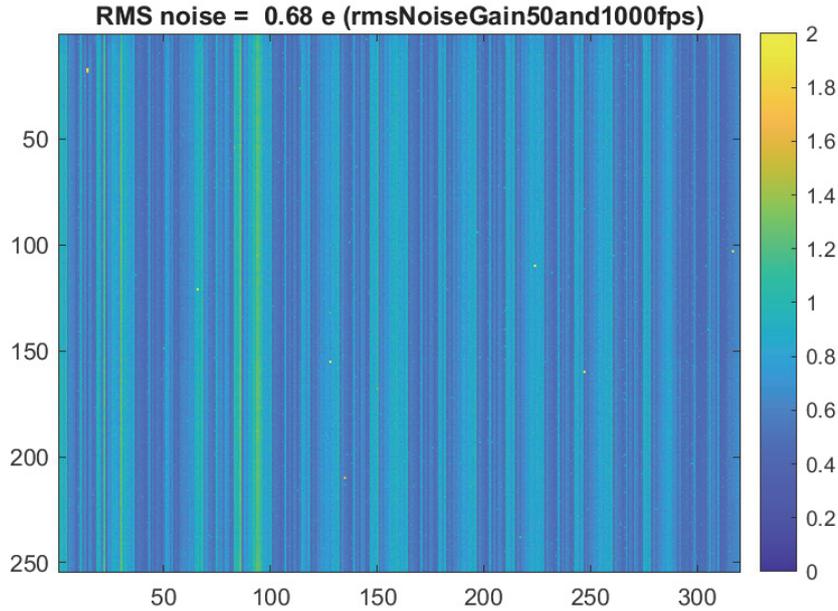

*Figure 7 : CDS noise, Gain x50, 1000 fps*

## 2.5 Dark current

To do this measurement the sensor is in the dark, looking at an 80K cold stop. The dark current is measured by fitting a line over the ADU level vs exposure time graph. The slope of this line gives the mean dark count. The measurements show that the dark current depends on the readout mode and the readout speed of the sensor as plotted in Figure 6. This is due to ROIC glowing.

Conditions of set-up:

- Detector temperature: 80K
- Read-out mode: GRS (Global Reset Single Readout)
- eAPD gain: 10
- Cold stop in front of the detector (inside the cryostat); cold baffle and filter are not mounted
- Water temperature: 20°C.

The integration time is varied by step from 1s to 100 s. For each integration time, 10 images are stored and averaged. The average signal is converted in e for each integration time and is plotted as a function of the integration time. The slope of the linear regression of this curve gives the dark in e/s/pixel.

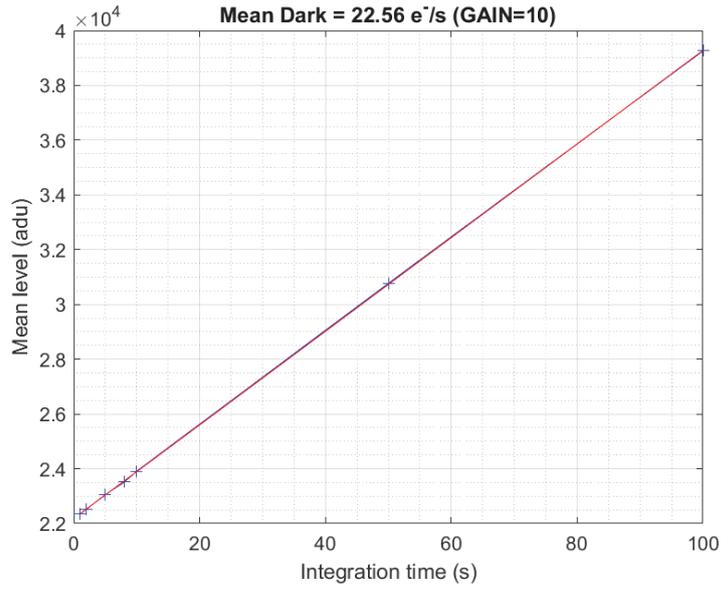
Figure 8 : mean dark current of the camera with an 80K cold stop

The Figure 8 shows a typical measurement of the dark current with a cold stop, with a value of 22.5 e/s.

**2.6 Background measurement**

The background current is measured the same way it is for the dark current but looking at a room temperature blackbody.

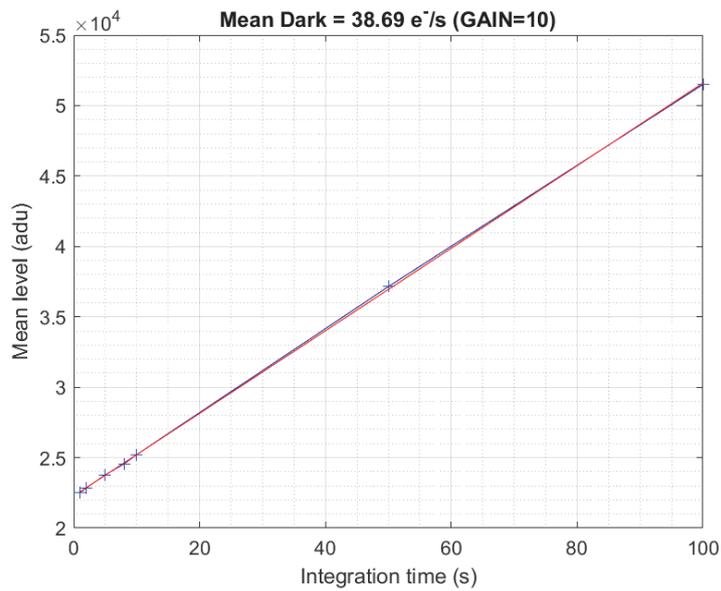
Figure 9 : mean background of the camera with a hot mirror in front of the camera window

The Figure 8 shows a typical measurement of the camera background, with a value of 38.7 e/s.

With the simultaneous improvement of the high wavelengths filtering and the use of ME1001 ROIC, the dark and the background were improved compared to the previous version of the camera. We measure typically in H band with F/4 aperture: 22 e/s dark and 38 e/s background at a temperature of 80K.

## 2.7 Cosmetics

One of the advantages of this sensor is its extremely good cosmetics, even when high gain is applied. Some other groups reported only a few dead pixels over the entire array, which is due to the HgCdTe growing process (MOVPE). We now proceed to bad pixels count of our camera. 0 bad pixel cameras are routinely delivered, which is unique for SWIR cameras.

The cosmetics specifications are given hereafter:

- Operability due to signal response

Pixels with signal < 0.5*median < 0.1 % at bias of 9V and integration time of 10 ms.

- Operability due to CDS noise

Pixels with noise > 2*median < 0.1 % at bias of 9V and integration time of 10 ms.

An example of 0 bad pixel count due to the signal is given in the Figure 10.

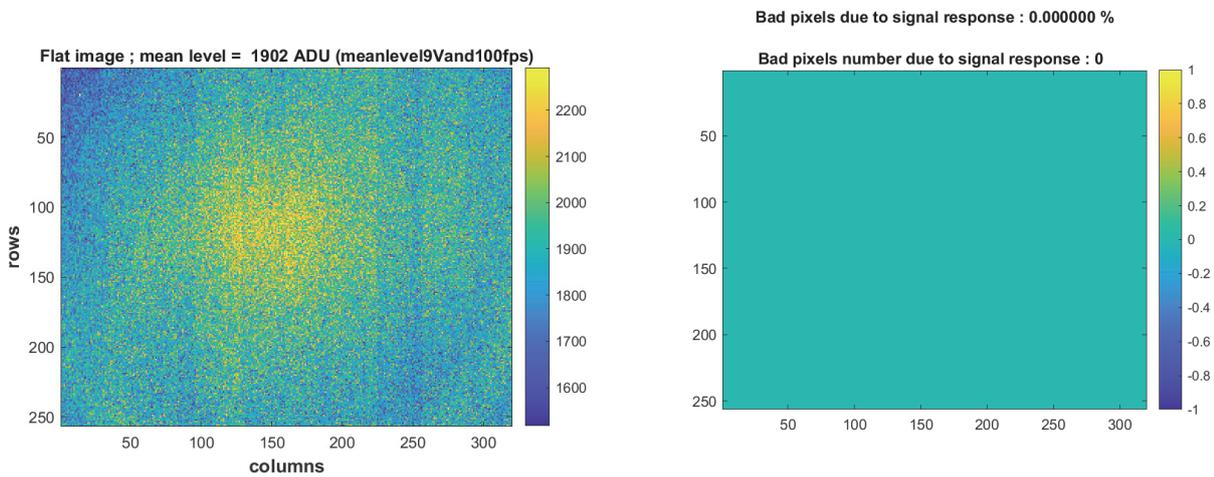

Figure 10 : (left) flat field image used for bad pixels count; (right) 0 bad pixel count due to signal response

## 2.8 Vibrations measurements

A special care was made to damp as much as possible the pulse tube vibrations. The mechanical design of the camera has been improved with a better mechanical isolation of the pulse tube from the focal plane.

A set-up was design to measure the vibrations of the detector with respect to the front flange of the camera.

Conditions of set-up:

- Detector temperature: 80K
- Read-out mode: CDS full frame 1750 fps
- eAPD gain: 1
- Cold baffle and filters are mounted

- Water temperature: 20°C.

A light source shall be imaged on the detector surface. The diameter of the spot imaged on the detector shall be at least 10 pixels and no more than 15 pixels. No pixel shall contain more than 10% of the total flux of the spot. The signal shall be around 15000 ADU/pixel in CDS mode. 2000 images shall be acquired. The centroid shall be calculated on each image after background subtraction. The standard deviation of the centroid shall be computed along the X and Y axes of the detector.

The Figure 11 shows the spot and the window used for the jitter measurement. The analysis is based on centroid measurement of the spot in the conditions detailed before. The window shown in this figure is also the window on which the centroid is computed.

For each image of the cube, the centroid is computed on the X and Y axis allowing to perform the required statistics on the X and Y jitter.

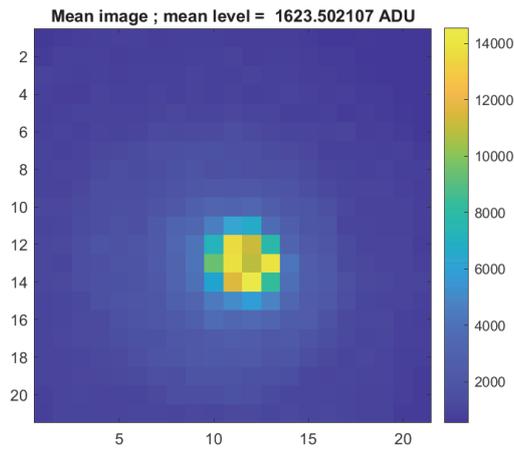

*Figure 11: spot and window used for the vibration measurements*

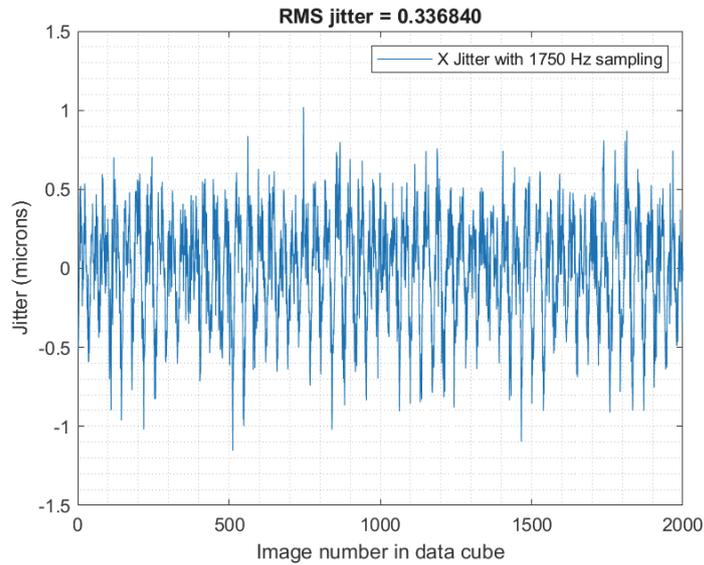

*Figure 12: X axis RMS jitter (in µm).*

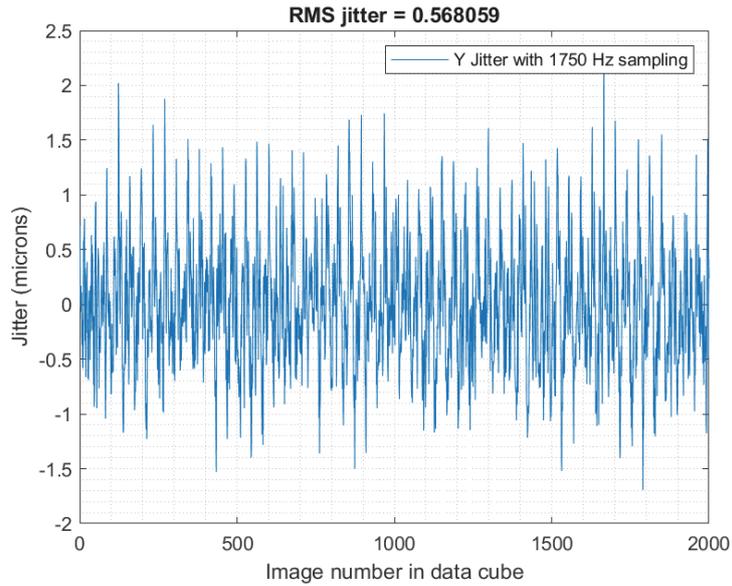

*Figure 13: Y axis RMS jitter (in μm).*

The Figure 12 and Figure 13 show the vibrations results for the X and Y axis. From these figures, we can see that vibrations are at the level of:

- **0.34 μm** RMS for the X axis
- **0.57 μm** RMS for the Y axis

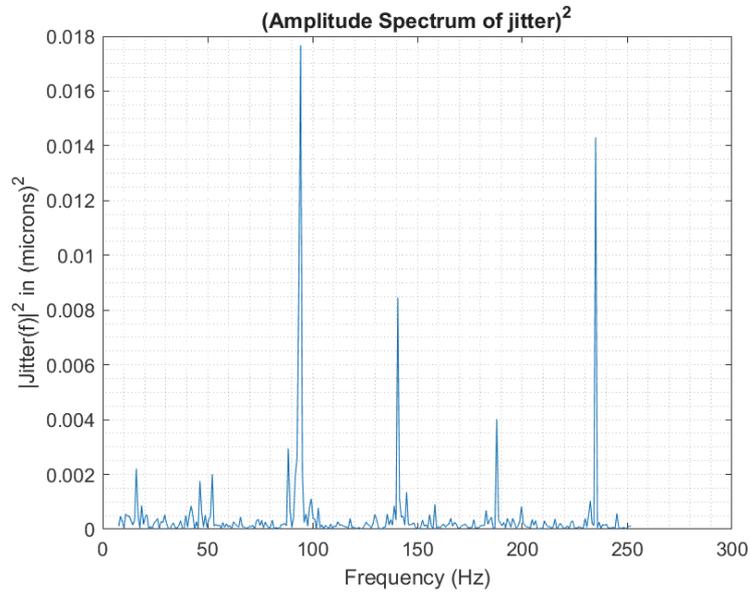

*Figure 14: FFT of the radial jitter for frequency analysis*

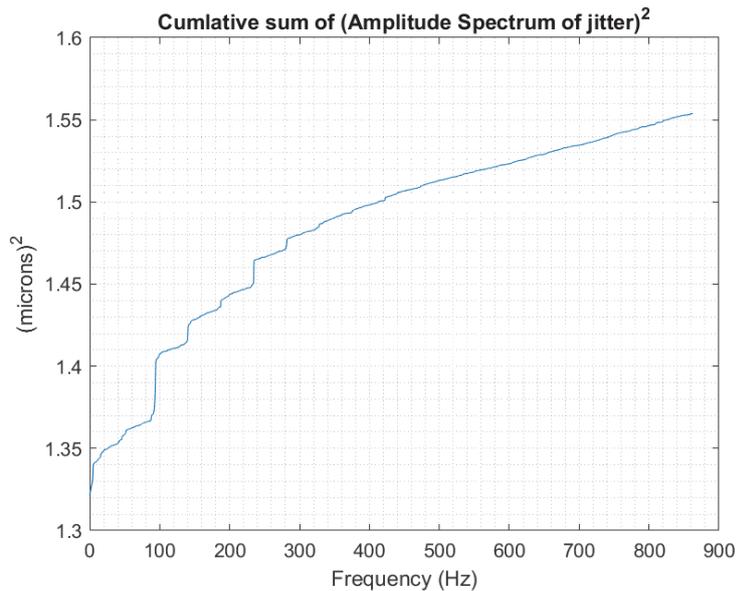

*Figure 15: cumulative sum of the radial jitter amplitude as a function of the frequency*

The main peaks that are shown in these figures are at 47 and 94 Hz and their harmonics. This corresponds to the pulse tube frequency and its harmonics, which is what we expected.

## 2.9 Suppression of the bias oscillation

With the first generation of C-RED One camera, the bias of the image was oscillating in phase with the pulse tube drive. An improved grounding scheme of the electronics improved deeply this issue.

The figure below clearly shows the issue. A set of 1000 images in non-destructive readout mode is recorded in the darkness. With the camera background, the signal increases with a linear shape on which a sinusoidal oscillation is added. On the left part of the figure, the oscillation in present. On the right part of the figure, we can see that the oscillation is damped.

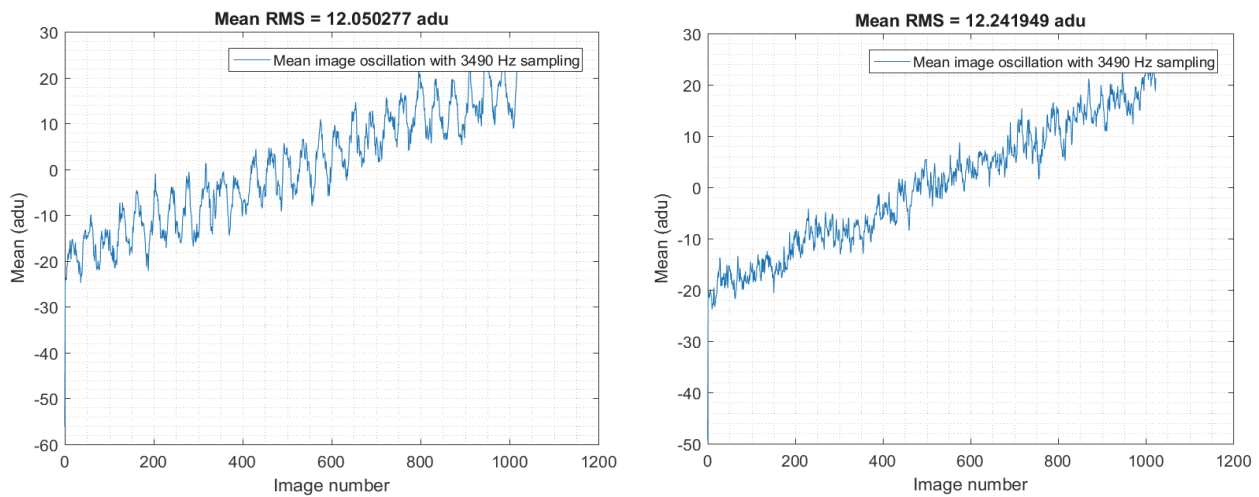

*Figure 16 : (left) oscillation of the mean bias value in the former version of C-RED One; (right) bias oscillation reduction with improved grounding.*

A frequency analysis of this oscillations allows to understand better the issue.

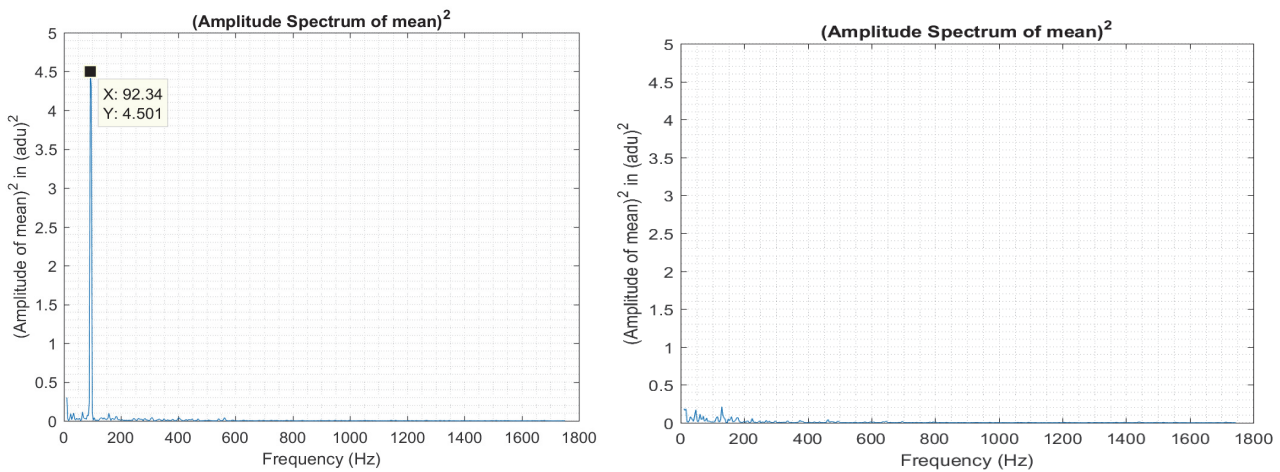

*Figure 17 : bias oscillation frequency analysis; (left) before bias oscillation removal; (right) after bias oscillation removal.*

On the left part of the Figure 17, the frequency analysis shows a clear peak at 92 Hz, which is in fact the double of the pulse tube frequency (46-47 Hz). On the right part of this figure, we can see that the peak is eliminated, confirming that the bias oscillation is removed by a better grounding scheme.

## 3. CONCLUSION

The C-RED One camera, developed by First Light Imaging for fast ultra-low noise infrared applications, is now a mature product with unprecedented performances. This camera uses the Leonardo Saphira e-APD 320x256 infrared sensor in an autonomous cryogenic environment with a low vibration pulse tube and with embedded readout electronics system. Several aspects of this camera were improved: the bias oscillation was removed, the detector vibrations are now damped at a level of 0.3 µm RMS jitter. The camera background was reduced by the use of the ME1001 ROIC removing the intrinsic glow source of the ROIC. The camera background in H band at 80K with a f/4 beam aperture is now of the level of 30-40 e/s at gain 10. At gain 50 and in the same conditions, total noise with a cap in front of the camera window is of the order of 0.7 e RMS. The cosmetic of the camera is measured in house. A camera with very few, or even 0, defective pixels are now delivered thanks to the Leonardo quality of the MCT growth by MOVPE. At this date, 40 C-RED One cameras have been delivered all over the world by First Light Imaging.